\documentclass[11pt]{llncs}

\usepackage{amsfonts,amsmath,amssymb,bm}

\newcommand{\inv}{^{-1}}
\newcommand{\orinv}{^{\pm1}}
\newcommand{\la}{\leftarrow}
\newcommand{\ra}{\rightarrow}
\newcommand{\lnf}{\ell_\text{NF}}
\newcommand{\dAs}{d_{A_s}}
\newcommand{\dAsW}{\overline{\dAs}}
\newcommand{\dAsM}{d^m_{A_s}}
\newcommand{\dBs}{d_{B_s}}
\newcommand{\dBsW}{\overline{\dBs}}
\newcommand{\nfi}{\textbf{(NF1)}}
\newcommand{\nfii}{\textbf{(NF2)}}

\newtheorem{algo}{Algorithm}
\newtheorem{defn}{Definition}
\newtheorem{remk}{Remark}
\newtheorem{clmm}{Claim}
\newtheorem{lemm}{Lemma}
\newtheorem{xmpl}{Example}

\begin{document}

\title{Cryptanalysis of group-based key agreement protocols using
subgroup distance functions}
\author{Dima Ruinskiy \and Adi Shamir \and Boaz Tsaban}
\institute{Weizmann Institute of Science, Rehovot, Israel}

\maketitle
\begin{abstract}
We introduce a new approach for cryptanalysis of key agreement
protocols based on noncommutative groups. This approach uses
functions that estimate the distance of a group element to a given
subgroup. We test it against the Shpilrain-Ushakov protocol, which
is based on Thompson's group $F$, and show that it can break about
half the keys within a few seconds on a single PC.
\\\\
\keywordname{\enspace Key agreement, Cryptanalysis, Thompson's
group, Shpilrain-Ushakov, Subgroup distance function}
\end{abstract}

\section{Introduction}

Key agreement protocols have been the subject of extensive studies
in the past 30 years. Their main task is to allow two parties (in
the sequel, Alice and Bob) to agree on a common secret key over an
insecure communication channel. The best known example of such a
protocol is the Diffie-Hellman protocol, which uses a (commutative)
cyclic group. Over the last few years, there was a lot of interest
in key agreement protocols based on noncommutative groups, and much
research was dedicated to analyzing these proposals and suggesting
alternative ones (see, e.g.,
\cite{AAG,Ts1,Ts2,HugTan,KoLee,Matucci,ShpAssessing,Unnec,ShpUsh},
and references therein).

A possible approach for attacking such systems is the
\emph{length-based cryptanalysis}, which was outlined in
\cite{HugTan}. This approach relies on the existence of a good
length function on the underlying group, i.e., a function $\ell(g)$
that tends to grow as the number of generators multiplied to obtain
$g$ grows. Examples of groups known to have such length functions
are the \emph{braid group $B_N$} \cite{Artin} and \emph{Thompson's
group $F$} \cite{IntroThompson}. For these groups, several practical
realizations of length-based attacks were demonstrated
\cite{Ts1,Ts2,self}. These attacks can achieve good success rates,
but usually only when we allow the algorithm to explore many
suboptimal partial solutions, which greatly increases both the time
and space complexities (see \cite{Ts2} for more details).

We introduce a novel approach to cryptanalysis of such key agreement
protocols, which relies on the notion of \emph{subgroup distance
functions}, i.e., functions that estimate, for an element $g \in G$
and a subgroup $H \le G$, the distance from $g$ to $H$. The
motivation for these distance-based attacks is the fact that several
families of public key agreement protocols suggest predefined pairs
of subgroups of the main group to be used for key generation, and
their security depends on the ability of the adversary to generate
any elements in these subgroups, which are in some way equivalent to
the originals (see \cite{self,Unnec}). We construct the theoretical
framework for \emph{distance-based attacks} and demonstrate its
applicability using the Shpilrain-Ushakov protocol in Thompson's
group $F$ \cite{ShpUsh} as an example. Although it has recently been
shown by Matucci \cite{Matucci} that the implementation of the
proposed protocol in $F$ can be broken deterministically using a
specialized attack based on the structural properties of the group,
it is still an interesting test case for more generic attacks, such
as the one proposed here.

The paper is organized as follows: in Section \ref{sec:proto} we
present the protocol in its general form. We then introduce in
Section \ref{sec:subgdist} the notion of subgroup distance function
and a general attack scheme based on it. Section \ref{sec:thomp}
describes the setting for the protocol in Thompson's group $F$. In
Section \ref{sec:thompdist} we introduce several subgroup distance
functions in $F$. Section \ref{sec:distatk} describes our
experimental cryptanalytic results.

\section{The Shpilrain-Ushakov Key agreement
Protocol}\label{sec:proto}

The protocol below was suggested by Shpilrain and Ushakov in
\cite{ShpUsh}. The authors suggested to use Thompson's group $F$ for
its implementation. Before we focus on that example, we'll discuss
the general case.

\begin{enumerate}
\item[(0)] Alice and Bob agree (publicly) on a group $G$ and
subgroups $A,B \le G$, such that $ab=ba$ for each $a \in A$ and each
$b \in B$.
\item A public word $z \in G$ is selected.
\item Alice selects privately at random
elements $a_1 \in A$ and $b_1 \in B$, computes $u_1 = a_1 z
b_1$, and sends $u_1$ to Bob.
\item Bob selects privately at random elements $a_2
\in A$ and $b_2 \in B$, computes $u_2 = b_2 z a_2$, and sends
$u_2$ to Alice.
\item Alice computes $K_A = a_1 u_2 b_1 = a_1 b_2 z a_2 b_1$,
whereas Bob computes $K_B = b_2 u_1 a_2 = b_2 a_1 z b_1 a_2$.
\end{enumerate}

As $a_1 b_2 = b_2 a_1$ and $a_2 b_1 = b_1 a_2$, $K_A=K_B=K$ and so
the parties share the same group element, from which a secret key
can be derived.

\subsection{Breaking the protocol}\label{subsec:breaking}

The goal of the adversary is to obtain the secret group element $K$
from the publicly known elements $u_1$, $u_2$ and $z$. For this it
suffices to solve the following problem:

\begin{defn}[Decomposition problem]
Given $z\in G$ and $u=azb$ where $a\in A$ and $b\in B$, find some
elements $\tilde a\in A$ and $\tilde b\in B$, such that $\tilde
az\tilde b=azb$.
\end{defn}

Indeed, assume that the attacker, given $u_1 = a_1 z b_1$, finds
$\tilde a_1 \in A$ and $\tilde b_1 \in B$, such that $\tilde a_1 z
\tilde b_1 = a_1 z b_1$. Then, because $u_2 = b_2 z a_2$ is known,
the attacker can compute $$\tilde a_1 u_2 \tilde b_1 = \tilde a_1
b_2 z a_2 \tilde b_1 = b_2 \tilde a_1 z \tilde b_1 a_2 = b_2 u_1 a_2
= K_B\enspace.$$ Alternatively, the attacker can break the protocol
by finding a valid decomposition of $u_2 = b_2 z a_2$.

For any given $\tilde a\in A$ we can compute its \emph{complement}
$\tilde b = z\inv \tilde a\inv u = z\inv \tilde a\inv(azb)$, which
guarantees that $\tilde a z\tilde b = azb$. The pair $\tilde
a,\tilde b$ is a solution to this problem if, and only if, $\tilde
b\in B$. A similar comment applies if we start with $\tilde b\in B$.
This involves being able to solve the \emph{group membership
problem}, i.e., to determine whether $\tilde b\in B$ (or $\tilde
a\in A$ in the second case).

It should be stressed that solving the decomposition problem is
sufficient, but not necessary in order to cryptanalyze the system.
All that is required in practice is finding some pair $\tilde
a,\tilde b$ that succeeds in decrypting the information passed
between Alice and Bob. Any pair $\tilde a\in A$ and $\tilde b\in B$
will work, but there can be other pairs, which are just as good.
This observation can be useful in cases where the group membership
problem is difficult or in groups where the centralizers of
individual elements are considerably larger than the centralizers of
the subgroups (which is not the case in $F$, see \cite{Thomp1}). For
simplicity, in the sequel we will restrict ourselves to solutions
where $\tilde a\in A$ and $\tilde b\in B$.

\section{Subgroup distance functions}\label{sec:subgdist}

\begin{defn}[Subgroup distance function]
Let $G$ be a group, $H \le G$ a subgroup. A function $d_H : G \to
\mathbb{R}^+$ is a \emph{subgroup distance function} if it satisfies
the following two axioms:
\begin{enumerate}
\item Validity: $d_H(h)=0$ for all $h \in H$.
\item Non-triviality: $d_H(g)>0$ for all $g \not\in H$.
\end{enumerate}
It is an \emph{invariant} subgroup distance function if it also
satisfies:
\begin{enumerate}
\item[(3)] Invariance: $d_H(gh) = d_H(hg) = d_H(g)$ for all $g \in G$ and
$h \in H$.
\end{enumerate}
\end{defn}

Clearly, if it is possible to evaluate a subgroup distance function
$d_H$ on all elements of $G$, then the membership decision problem
for $H$ is solvable: $g \in H \iff d_H(g)=0$. Conversely, if one can
solve the membership decision problem, a trivial distance function
can be derived from it, e.g., $d_H(g)=1-\chi_H(g)$, where $\chi_H$
is the characteristic function of $H$.

Obviously, this trivial distance function is not a good example. For
the subgroup distance function to be useful, it has to somehow
measure how close a given element $g$ is to $H$, that is, if
$d_H(g_1)<d_H(g_2)$, then $g_1$ is closer to $H$ than $g_2$. This
concept of ``closeness'' can be hard to define, and even harder to
evaluate. The notion of what's considered a good distance function
may vary, depending on the subgroups and on the presentation. In the
sequel we will discuss concrete examples of subgroup distance
function in Thompson's group $F$.

Assuming the existence of such functions, consider the following
algorithm for solving the decomposition problem:

\begin{algo}[Subgroup distance attack]\label{algo:dist}
~\\\\
We are given words $z,xzy \in G$, where $x \in X$ and $y \in Y$,
$X,Y$ are commuting subgroups of $G$ and $S_X,S_Y$ are their
respective (finite) generating sets. The goal it to find some
$\tilde x \in X$ and $\tilde y \in Y$, such that $xzy=\tilde x z
\tilde y$. The algorithm runs at most a predefined number of
iterations $N$.
\begin{enumerate}
\item Let $\tilde x \la 1$.
\item\label{itm:step} For each $g_i \in S_X\orinv$ compute $x_i =
\tilde x g_i$, its complement $y_i = z\inv x_i\inv xzy$ and
evaluate $d_Y(y_i)$. If $d_Y(y_i)=0$, let $\tilde x=x_i$,
$\tilde y=y_i$ and halt.
\item Let $j$ be the index of the minimum $d_Y(y_i)$ (if several such $j$
are possible, choose one arbitrarily).
\item If the maximal number of iterations $N$ has been reached,
terminate. Otherwise, let $\tilde x \la x_j$ and return to step
\ref{itm:step}.
\end{enumerate}
\end{algo}

Observe that if the algorithm halts in step \ref{itm:step}, then the
pair $\tilde x, \tilde y$ is a solution of the decomposition
problem.

Algorithm \ref{algo:dist} is very similar to the length-based
attacks described in \cite{Ts1,self}. The difference is that it uses
the subgroup distance function, instead of the length function to
evaluate the quality of candidates. As such, any extensions
applicable to the length-based algorithms (such as memory,
lookahead, etc.) can be used with the distance-based attack as well.
Refer to \cite{Ts2,self} for more information.

\subsection{Attacking the Shpilrain-Ushakov
protocol}

The adversary is given the common word $z$ and the public elements
$u_1,u_2$. These can be translated into four equations in the group:

\begin{equation}\label{eq:4eq}
\begin{array}{rcl}
u_1 & = & a_1zb_1\\
u_2 & = & b_2za_2\\
u_1\inv & = & b_1\inv z\inv a_1\inv\\
u_2\inv & = & a_2\inv z\inv b_2\inv
\end{array}
\end{equation}

Algorithm \ref{algo:dist} (with or without possible extensions) can
be applied to each of the four equations separately, thus attacking
each of the four private elements $a_1,a_2,b_1\inv,b_2\inv$. A
single success out of the four attempts is sufficient to break the
cryptosystem (see Section \ref{subsec:breaking}).

\section{Thompson's group}\label{sec:thomp}

Thompson's group $F$ is the infinite noncommutative group defined by
the following generators and relations:
\begin{equation}\label{eq:F}
F = \langle \ x_0 , x_1 , x_2 , \dots \ | \ x_i\inv x_k x_i =
x_{k+1} \ (k>i) \ \rangle
\end{equation}

\begin{remk}\label{remk:pairgen}
From Equation \eqref{eq:F} it's evident that the elements $x_0,x_1$
and their inverses generate the entire group, because $x_k\orinv =
x_0^{1-k} x_1\orinv x_0^{k-1}$ for every $k\ge2$.
\end{remk}

\begin{defn}
A basic generator $x_i\orinv$ of $F$ is called a \emph{letter}. A
generator $x_i$ is a \emph{positive letter}. An inverse $x_i\inv$ is
a \emph{negative letter}. A \emph{word} in $F$ is a sequence of
letters. We define $|w|$ as the length of the word $w$ , i.e., the
number of letters in it.
\end{defn}

\begin{defn}
A word $w \in F$ is said to be in \emph{normal form}, if
\begin{equation}
w = x_{i_1} \cdots x_{i_r} x_{j_t}\inv \cdots x_{j_1} \inv
\end{equation}
and the following two conditions hold:
\begin{enumerate}
\item[\nfi] $i_1 \leq\dots\leq i_r$ and $j_1 \leq\dots\leq j_t$
\item[\nfii] If both $x_i,x_i\inv$ occur in $w$, then at least one of
$x_{i+1},x_{i+1}\inv$ occurs too.
\end{enumerate}
A word is said to be in \emph{seminormal form} if only \nfi\  holds.
\end{defn}

While a seminormal form is not necessarily unique, a normal form is,
i.e., two words represent the same group element if and only if they
have the same normal form \cite{IntroThompson}. The following
rewriting rules can be used to convert any word to its seminormal
form \cite{ShpUsh}:

For all non-negative integers $i<k$:
\begin{displaymath}
\begin{array}{llcl}
(R1) & x_k x_i & \ra & x_i x_{k+1} \\
(R2) & x_k\inv x_i & \ra & x_i x_{k+1}\inv \\
(R3) & x_i\inv x_k & \ra & x_{k+1} x_i\inv \\
(R4) & x_i\inv x_k\inv & \ra & x_{k+1}\inv x_i\inv \\
\end{array}
\end{displaymath}

For all non-negative integers $i$:
\begin{displaymath}
\begin{array}{llcl}
(R5) & x_i\inv x_i\phantom{\inv} & \ra & 1\phantom{x_ix_{k+1}\inv} \\
\end{array}
\end{displaymath}

The seminormal form can be subsequently converted to a normal form
by searching for pairs of indices violating \textrm{(NF2)}, starting
from the boundary between the positive and negative parts, and
applying the inverses of rewriting rules (R1) and (R4) to eliminate
these pairs \cite{ShpUsh}:

Suppose that $(x_{i_a},x_{j_b}\inv)$ is a pair of letters violating
\textrm{(NF2)} and that $a$ and $b$ are maximal with this
property (i.e., there exists no violating pair
$(x_{i_k},x_{j_l}\inv)$ with $k>a$ and $l>b$). Then $i_a=j_b$ and
all indices in $x_{i_{a+1}} \cdots x_{i_r} x_{j_t}\inv \cdots
x_{j_{b+1}}\inv$ are higher than $i_a+1$ (by definition of
\textrm{(NF2)}). Applying the inverse of (R1) to $x_{i_a}$ and the
inverse of (R4) to $x_{j_b}\inv$ we get:

\begin{displaymath}
\begin{array}{lcl}
w & = & x_{i_1} \cdots x_{i_a} \underbrace{(x_{i_{a+ 1}} \cdots
x_{i_r} x_{j_t}\inv \cdots x_{j_{b+1}}\inv)}_{c} x_{j_b}\inv \cdots
x_{j_1} \\ & \ra & x_{i_1} \cdots x_{i_{a+1}-1} \cdots x_{i_r-1}
\underbrace{(x_{i_a} x_{j_b}\inv)}_{cancel} x_{j_t-1}\inv \cdots
x_{j_{b+1}-1}\inv \cdots x_{j_1} \\ & \ra & x_{i_1} \cdots
x_{i_{a-1}} \underbrace{(x_{i_{a+1}-1} \cdots x_{i_r-1}
x_{j_t-1}\inv \cdots x_{j_{b+1}-1}\inv)}_{c'} x_{j_{b-1}} \cdots
x_{j_1}
\end{array}
\end{displaymath}

The violating pair $(x_{i_a},x_{j_b}\inv)$ is cancelled and the
subword $c'$ obtained from $c$ by index shifting contains no
violating pairs (by the assumption of maximality on $(a,b)$). Thus,
we can continue searching for bad pairs, starting from $a-1$ and
$b-1$ down. Thus we are guaranteed to find and remove all the
violating pairs and reach the normal form.

\begin{defn}[Normal form length]
For $w \in F$, whose normal form is  $\hat w$, define the
\emph{normal form length} as $\lnf(w)=|\hat w|$.
\end{defn}

The following lemma shows the effect multiplication by a single
letter has on the normal form of the word. This result will be
useful in the following sections.

\begin{lemm}\label{lemma:pm1}
Let $w \in F$ and $x=x_t\orinv$ be a basic generator of $F$ in the
presentation \eqref{eq:F}. Then $\lnf(xw)=\lnf(w)\pm1$ (and due to
symmetry, $\lnf(wx)=\lnf(w)\pm1)$.
\begin{proof}
We'll concentrate on the product $xw$ (obviously, the case of $wx$
is similar) and observe what happens to the normal form of $w$ when
it's multiplied on the left by the letter $x$. Without loss of
generality, $w = x_{i_1} \cdots x_{i_k} x_{j_l}\inv \cdots x_{j_1}
\inv$ is in normal form. Denote the positive and negative parts of
$w$ by $w_p$ and $w_n$ respectively.

Assume that $x=x_t$ is a positive letter. Then $bw$ is converted to
a seminormal form by moving $x$ into its proper location, while
updating its index, using repeated applications of (R1). Assuming
$m$ applications of (R1) are necessary, the result is of the form:
$$\overline{bw} = x_{i_1} \cdots x_{i_m} \bm{x_{t+m}} x_{i_{m+1}} \cdots
x_{i_k} x_{j_l}\inv \cdots x_{j_1}\inv\enspace,$$ where $i_m <
t+m-1$ and $i_{m+1} \ge t+m$.

\begin{remk}\label{remk:ast}
Observe that it is not possible that $i_m=t+m-1$, because in order
to apply (R1): $x_{t+m-1}x_{i_m} \ra x_{i_m}x_{t+m}$, one must have
$i_m < t+m-1$.
\end{remk}

\begin{xmpl}
$w=x_3x_7x_{11}x_9\inv x_4\inv$, $b=x_8$. $bw=x_8 \cdot
x_3x_7x_{11}x_9\inv x_4\inv$ is converted to
$\overline{bw}=x_3x_7\bm{x_{10}}x_{11}x_9\inv x_4\inv$, by 2
applications of (R1).
\end{xmpl}

Obviously, $\overline{bw}$ is a seminormal form and
$|\overline{bw}|=|w|+1$. If $\overline{bw}$ is in normal form (as in
the above example), we're done. The only situation where it's not in
normal form, is if it contains pairs violating \nfii. Since
$x_{t+m}$ is the only letter introduced, the only violating pair can
be $(x_{t+m},x_{t+m}\inv)$. This may occur, if $w$ contained
$x_{t+m}\inv$, but neither $x_{t+m}$, nor $x_{t+m+1}\orinv$.

\begin{xmpl}
$w=x_3x_7x_{11}x_9\inv x_4\inv$, $b=x_7$. $bw=x_7 \cdot
x_3x_7x_{11}x_9\inv x_4\inv$ is converted to
$\overline{bw}=x_3x_7\bm{x_9}x_{11}x_9\inv x_4\inv$. In this case
$(x_9,x_9\inv$) violates \nfii. The inverse of (R1) is applied to
rewrite $x_9x_{11} \ra x_{10}x_9$, and $x_9x_9\inv$ are canceled
out, yielding the (normal) word $\widehat{bw}=x_3x_7x_{10}x_4\inv$.
\end{xmpl}

Whenever a situation occurs as described above, the pair
$(x_{t+m},\allowbreak x_{t+m}\inv)$ is cancelled, according to the
procedure described in Section \ref{sec:thomp}. This causes all
indices above $t+m$ to be decreased by 1. The resulting word is
$$\widehat{bw} = x_{i_1} \cdots x_{i_m} x_{i_{m+1}-1} \cdots
x_{i_k-1} x_{j_l-1}\inv \cdots x_{j_{n+1}-1}\inv x_{j_n}\inv \cdots
x_{j_1}\inv\enspace,$$ where $i_m < t+m-1$, $i_{m+1} \ge t+m+2$,
$j_n \le t+m$ and $j_{n+1} \ge t+m+2$. We have
$|\widehat{bw}|=|w|-1$ and, in fact, $\widehat{bw}$ is in normal
form. Indeed, once the pair $(x_{t+m},x_{t+m}\inv)$ is cancelled,
the only new pair violating \nfii\ that can be introduced is
$(x_{t+m-1},x_{t+m-1}\inv)$, but this is not possible, because
$x_{t+m-1}$ does not appear in $\widehat{bw}$, due to Remark
\ref{remk:ast}. This completes the proof for positive letters.

Now, consider the case where $x=x_t\inv$, a negative letter. $bw$ is
converted to a seminormal form by moving $x_t\inv$ to the right,
while updating its index, using the different rewriting rules. There
are two possible outcomes:

\textbf{(1)} After $m$ applications of (R2) the resulting word is
$$\overline{bw} = x_{i_1} \cdots x_{i_m} \bm{x_{t+m}\inv}
x_{i_{m+1}} \cdots x_{i_k} x_{j_l}\inv \cdots x_{j_1}\inv\enspace,$$
where $i_{m+1}=t+m$, and so the pair is cancelled by applying (R5).
Now, because $i_m<t+m-1$, the elimination of the pair
$(x_{t+m},x_{t+m}\inv)$ does not introduce pairs that violate \nfii,
and so $\overline{bw}$ is in normal form and has
$|\overline{bw}|=|w|-1$.

\begin{xmpl}
$w=x_3 x_7 x_9\inv x_4\inv$, $b=x_6\inv$. $bw=\bm{x_6\inv} x_3 x_7
x_9\inv x_4\inv$ is converted to  $x_3 \bm{x_7\inv x_7} x_9\inv$ and
the pair of inverses is cancelled out to obtain $x_4\inv \ra x_3
x_9\inv x_4\inv$.
\end{xmpl}

\textbf{(2)} $x_t\inv$ is moved to its proper place among the
negative letters, updating its index if necessary. This is completed
through $m$ applications of (R2), followed by $k-m$ applications of
(R3) and finally, $l-n$ applications of (R4), to obtain
$$\overline{bw} = x_{i_1} \cdots x_{i_m} \bm{x_{t+m}\inv}
x_{i_{m+1}+1} \cdots x_{i_k+1} x_{j_l+1}\inv \cdots
x_{j_{n+1}+1}\inv \bm{x_{t+m}\inv} x_{j_n}\inv \cdots
x_{j_1}\inv\enspace,$$ where $i_m<t+m-1$, $i_{m+1}>t+m$,
$j_{n+1}>t+m$ and $j_n \le t+m$. Because the letter $x_{t+m}$ is not
present in $\overline{bw}$ (otherwise the previously described
situation would occur), the newly introduced letter $x_{t+m}\inv$
cannot violate \nfii, and therefore $\overline{bw}$ is in fact in
normal form and $|\overline{bw}|=|w|+1$.

\begin{xmpl}
$w=x_3 x_7 x_9\inv x_4\inv$, $b=x_5\inv$. $bw=\bm{x_5\inv} x_3 x_7
x_9\inv x_4\inv$ is rewritten as: $\bm{x_6\inv} x_7 x_9\inv x_4\inv
\ra x_3 x_8 \bm{x_6\inv} x_9\inv x_4\inv \ra x_3 x_8 x_{10}\inv
\bm{x_6\inv} x_4\inv$.
\end{xmpl}

\noindent This completes the proof for negative letters.

\qed\end{proof}
\end{lemm}

\subsection{The Shpilrain-Ushakov protocol in Thompson's
group}\label{subsec:param}

For a natural number $s \ge 2$ let $S_A = \{x_0 x_1\inv , \dots ,
x_0 x_s\inv\}$, $S_B = \{x_{s+1}, \dots ,\allowbreak x_{2s}\}$ and
$S_W = \{x_0, \dots , x_{s+2}\}$. $S_W$ generates $F$ (see Remark
\ref{remk:pairgen}). Denote by $A_s$ and $B_s$ the subgroups of $F$
generated by $S_A$ and $S_B$, respectively.

All of the following facts are shown in \cite{ShpUsh}: $A_s$ is
exactly the set of elements whose normal form is $$x_{i_1} \cdots
x_{i_m} x_{j_m}\inv \cdots x_{j_1}\inv\enspace,$$ i.e, has positive
and negative parts of the same length $m$, and additionally
satisfies $i_k-k<s$ and $j_k-k<s$ for every $k=1,\dots,m$. $B_s$ is
the set of all elements of $F$ whose normal form consists only of
letters with indices $\ge s+1$. Additionally, $A_s$ and $B_s$
commute elementwise, which makes them usable for implementing the
protocol in Section \ref{sec:proto}.

\subsubsection*{Key generation}

Let $s\ge2$ and $L$ be positive integers. The words $a_1, a_2\in
A_s$, $b_1, b_2\in B_s$, and $w\in F$ are all chosen of normal form
length $L$, as follows: Let $X$ be $A$, $B$, or $W$. Start with the
empty word, and multiply it on the right by a generator (or inverse)
selected uniformly at random from the set $S_X$. Continue this
procedure until the normal form of the word has length $L$.

For practical and (hopefully) secure implementation of the protocol,
it is suggested in \cite{ShpUsh} to use $s\in\{3,4,\dots,8\}$ and
$L\in\{256,258,\dots,320\}$.

\section{Subgroup distance functions in Thompson's
group}\label{sec:thompdist}

In this section we'll suggest several natural distance functions
from the subgroups $A_s,B_s \le F$ defined in Section
\ref{subsec:param}. These distance functions can be used to
implement the attack outlined by Algorithm \ref{algo:dist}.

\subsection{Distance functions from $B_s$}

For $w \in F$ define $P_i(w)$ and $N_i(w)$ as the number of
occurrences of $x_i$ and $x_i\inv$ in the normal form $\hat w$ of
$w$.

\begin{defn}[Distance from $B_s$]
Let $s\le2$ be an integer. For $w \in F$ the distance from $B_s$ is
defined as
$$\dBs(w) = \sum_{i=0}^s{\left(P_i(w)+N_i(w)\right)}$$
\end{defn}

\begin{clmm}
$\dBs$ is a distance function.
\begin{proof}
This is immediate, since an element is in $B_s$ if and only if its
normal form does not contain generators with indices below $s+1$
(see Section \ref{subsec:param}). \qed\end{proof}
\end{clmm}

\begin{clmm}\label{cl:binv}
$\dBs$ is an invariant distance function.

\begin{proof}

It is enough to consider only the generators of $B_s$. Indeed, if
multiplication by a single generator of $B_s$ does not change the
distance of a word $w$, neither does multiplication by a sequence of
these generators.

Let $w \in F$. Let $b=x_{s+\alpha}\orinv$, where $\alpha>0$. By
Lemma \ref{lemma:pm1}, we know that $b$ is either moved to its
proper position (and $\lnf(bw)=\lnf(w)+1$) or it is cancelled with
its inverse, either by (R5) or as part of a pair violating \nfii, in
which case $\lnf(bw)=\lnf(w)-1$. The index of $b$ is initially above
$s$, and may only increase when the rewriting rules are applied.
Therefore, if $b$ is cancelled at some point, the index of its
inverse is also above $s$. Furthermore, when pairs of elements are
rewritten, the lower-indexed element is not affected, so any letters
with indices $\le s$ will not be affected by moving $b$. Finally, if
$b$ is cancelled out due to violating \nfii, the process again only
affects letters with indices higher than $b$'s (see the proof of
Lemma \ref{lemma:pm1}). In all cases, the generators with indices
$\le s$ are not affected at all, and so $\dBs(bw)=\dBs(w)$.

\qed\end{proof}
\end{clmm}

One can intuitively feel that $\dBs$ is a natural distance function,
because it counts the number of ``bad'' letters in $w$ (letters that
do not belong to the subgroup $B_s$). Indeed, if $w$ is in normal
form, $w=w_pw_cw_n$, where $w_p$ and $w_n$ are the ``bad'' positive
and negative subwords, respectively, then $\dBs(w)=|w_p|+|w_n|$ and
$w_p\inv w w_n\inv \in B$.

We now introduce another natural function that measures distance
from $B_s$.

\begin{defn}[Weighted distance from $B_s$]
Let $s\le2$ be an integer. For $w \in F$ the weighted distance from
$B_s$ is defined as
$$\dBsW(w) = \sum_{i=0}^s{(s+1-i)\left(P_i(\hat w)+N_i(\hat w)\right)}$$
\end{defn}

$\dBsW$ does not only count the ``bad'' letters, but assigns a score
for each letter, depending on how far below $s+1$ it is (in
particular, $\dBs(w)\le\dBsW(w)$ for all $w \in F$. The following
claim is straightforward.

\begin{clmm}
$\dBsW$ is an invariant distance function.
\begin{proof}
The proof of Claim \ref{cl:binv} shows that multiplication by $b$
does not alter any letters below $s+1$ in $w$. Therefore, the weight
of each such letter is also preserved. \qed\end{proof}
\end{clmm}

\subsection{Distance functions from $A_s$}

We will now describe a number of natural distance functions from the
subgroup $A_s$. Recall (Section \ref{subsec:param}) that $A_s$ is
the set of all elements in $F$, whose normal form is of the type
$x_{i_1} \cdots x_{i_m} x_{j_m}\inv \cdots x_{j_1}\inv$, i.e, has
positive and negative parts of the same length $m$, and additionally
satisfies $i_k-k<s$ and $j_k-k<s$ for every $k=1,\dots,m.$

\begin{defn}[Distance from $A_s$]
Let $s\ge2$ be an integer. Let $w \in F$, such that its normal form
is $\hat w = x_{i_1} \cdots x_{i_p} x_{j_n}\inv \cdots x_{j_1}\inv$.
The distance from $A_s$ is defined as
$$\dAs(w) = \left|\{k:i_k-k \ge s\}\right|\ +\  \left|\{l
:j_l-l \ge s\}\right|\ +\ \left|p-n\right|$$
\end{defn}

$\dAs(w)$ is the number of ``bad'' letters in $\hat w$, i.e.,
letters that violate the $A_s$ property, plus the difference between
the lengths of the positive or negative parts. $\dAs$ is clearly a
distance function. However, it is not invariant, as shown by the
following example:

Similarly we can define a weighted distance function from $A_s$,
which not only counts the number of bad letters, but gives a score
to each such letter, based on the difference $i_k-k$ (or $j_k-k$).

\begin{defn}[Weighted distance from $A_s$]
Let $s\ge2$ be an integer. Let $w \in F$, such that its normal form
is $\hat w = x_{i_1} \cdots x_{i_p} x_{j_n}\inv \cdots x_{j_1}\inv$.
The weighted distance from $A_s$ is defined as
$$\dAsW(w) = \sum_{k=1 \dots
p}^{i_k-k \ge s}{(i_k-k-s+1)}\  + \sum_{k=1 \dots n}^{j_k-k \ge
s}{(j_k-k-s+1)}\ +\ \left|p-n\right|$$
\end{defn}

For each bad letter $x_{i_k}$ or $x_{j_k}\inv$, $\dAsW$ adds a
positive integer. As such, it's a distance function, which is again
not invariant (the example above works here too).

A somewhat different approach to defining distance from $A_s$ arises
from the observation that the number of bad letters can be less
important than the maximum value of the differences $i_k-k$ and
$j_k-k$ across the word, which measures the size of the violation.
The difference between the two distance functions roughly
corresponds to the difference between the $L_1$ and $L_\infty$
norms.

Let $\hat w = x_{i_1} \cdots x_{i_p} x_{j_n}\inv \cdots
x_{j_1}\inv$. Suppose that for some integer $k$ we have
$i_k-k-s+1=m_p > 0$ and that $m_p$ is the maximum for all $i_k$. By
multiplying the word by $x_0^{m_p}$ we shift the position for all
the original positive letters of $w$ by $m_p$, and so all of the
positive letters, including the first $m$ $x_0$'s have $i_k-k<s$.
Similarly, if $m_n$ is the maximum violation in the negative
subword, multiplication by $x_0^{-m_n}$ on the right eliminates all
violations among negative letters. However, this still does not mean
that the word is in $A_s$, because the positive and negative lengths
may differ. Let $\hat{w'}$ be the normal form obtained from
$\hat{w}$ through multiplication by $x_0^{m_p}$ and $x_0^{-m_n}$ on
the left and right, respectively. Let $l_p$ and $l_n$ be the
corresponding lengths of the positive and negative parts of
$\hat{w'}$. If $l_p-l_n>0$, then $\hat{w'}x_0^{l_n-l_p} \in A_s$. If
$l_p-l_n<0$, then $x_0^{l_n-l_p}\hat{w'} \in A_s$. Altogether, any
word can be changed to a word in $A_s$ through multiplication by
$m_p+m_n+|l_p+l_n|$ indices (when $l_p$ and $l_n$ are evaluated
\emph{after} multiplying by $x_0^{m_p}$ and $x_0^{-m_n}$).

This observation suggests the following distance function:

\begin{defn}[Maximum-based distance from $A_s$]
Let $s\ge2$ be an integer. Let $w \in F$, such that its normal form
is $\hat w = x_{i_1} \cdots x_{i_p} \allowbreak x_{j_n}\inv \cdots
x_{j_1}\inv$. Let $$m_p=\max{\left(\{0\}\cup\{i_k-k-s+1:k=1 \dots
p\}\right)}$$ and $$m_n=\max{\left(\{0\}\cup\{j_k-k-s+1:k=1 \dots
n\}\right)}\enspace.$$ The maximum-based distance from $A_s$ is
defined as
$$\dAsM(w)=m_p+m_n+\left|(p+m_p)-(n+m_n)\right|$$
\end{defn}

For every $w \in A_s$ $m_p$, $m_n$ and $|p-n|$ are $0$ by
definition, while for every $w \not\in A_s$ at least one of them has
to be positive, so the $\dAsM$ is a distance function. It turns out
that, unlike the two previously defined distance functions, $\dAsM$
is also invariant.

\begin{clmm}
$\dAsM$ is an invariant distance function.
\begin{proof}
As with Claim \ref{cl:binv}, it's sufficient to prove that
multiplication by a single generator of $A_s$ does not change the
distance from any word $w$ to $A_s$. We will consider
multiplications on the left by generators and their inverses. The
multiplication on the right follows symmetrically.

Let $w = x_{i_1} \cdots x_{i_p} x_{j_n}\inv \cdots x_{j_1}\inv$,
without loss of generality, in normal form. Consider the generator
$x_0x_t\inv$, where $1 \le t \le s$. Define $w'$ as the normal form
of $x_0x_t\inv w$. For the parameters $p,n,m_p,m_n$ of $w$, denote
by $p',n',m_p',m_n'$ their corresponding values in $w'$.

From Lemma \ref{lemma:pm1} it follows that each of the letters
$x_t\inv$ and $x_0$ can either be cancelled out with the appropriate
inverse, decreasing the length by 1, or placed in its appropriate
location, increasing the length by 1. There is a total of 4 possible
options:

\textbf{(1)} $x_t\inv$ is cancelled out, but $x_0$ is not: $w'=x_0
x_{i_1} \cdots x_{i_m} x_{i_{m+2}} \cdots \allowbreak x_{i_p}
x_{j_n}\inv \cdots x_{j_1}\inv$, where $x_{t+m}\inv$ is cancelled
out with $x_{i_{m+1}}$ after $m$ applications of (R2). It follows
that $p'=p$, $n'=n$ and $m_n'=m_n$ (because the negative letters are
unaffected). Observe also that there can be no bad letters among the
first $m$: indeed, (R2) is applied $m$ times, for each $k=1 \dots m$
rewriting $x_{t+k-1}\inv x_{i_k} \ra x_{i_k} x_{t+k}\inv$, so
necessarily $i_k < t+k-1$ for all $k$, or equivalently,
$i_k-k<t-1<s$. The multiplication by $x_0$ on the left only
increases their relative positions, thus decreasing $i_k-k$. Now,
any possible bad letters above $i_m$ are unchanged, and neither is
their relative position, so $m_p'=m_p$ and overall
$\dAsM(w')=\dAsM(w)$.

\textbf{(2)} Both $x_t\inv$ and $x_0$ are cancelled out:
$w'=x_{i_1-1} \cdots x_{i_m-1} x_{i_{m+2}-1} \cdots \allowbreak
x_{i_p-1} x_{j_n-1}\inv \cdots x_{j_{q+1}-1}\inv x_0^{1-q}$. Here
$p'=p-1$, $n'=n-1$ and $m_n'=m_n$ because all negative letters
$x_{j_k}\inv$ with $j_k>0$ had both their indices and their relative
positions decreased by 1. The same thing applies to positive letters
above $i_m$, which are the only positive letters that may be bad. So
again, $m_p'=m_p$ and $\dAsM(w')=\dAsM(w)$.

\textbf{(3)} Neither $x_t\inv$, nor $x_0$ are cancelled out: $w'=x_0
x_{i_1} \cdots x_{i_m} x_{i_{m+1}+1} \cdots \allowbreak x_{i_p+1}
x_{j_n+1}\inv \cdots x_{j_{q+1}+1}\inv \bm{x_{t+m}\inv} x_{j_q}\inv
\cdots x_{j_1}\inv$. Here $p'=p+1$ and $n'=n+1$. Due to the former
observation, bad positive letters may only exist beyond the first
$m$. All these letters had their indices $i_k$ and their relative
positions $k$ increased by 1, so the difference is preserved and
$m_p'=m_p$. Among the negative letters, only the letters whose
indices increased, also had their relative position increased, so
$j_k-k$ is preserved for all the original letters of $w$. Hence,
$m_n' \ge m_n$ and the only situation when it may actually increase
is when the new maximum is attained at the new letter, i.e.,
$m_n'=(t+m)-(q+1)-s+1 > m_n$. Because $t \le s$, $m \le p$ and $q
\le n$, we have $m_n' \le p-q$, from which it follows that
$$(p'+m_p')-(n'+m_n') = (p'-n') + (m_p'-m_n') = (p+1)-(n+1) + m_p -
m_n' \ge$$
$$\ge m_p + (p-n) - (p-q) = m_p+q-n \ge 0$$ Assuming
$m_n'>m_n$, it's obvious that
$$(p-n)+(m_p-m_n)>(p'-n')+(m_p'-m_n')\ge0\enspace,$$ and so if $m_n$
increases, $\left|(p+m_p)-(n+m_n)\right|$ decreases by the same
amount, and overall $\dAsM(w')=\dAsM(w)$.

\textbf{(4)} $x_t\inv$ is not cancelled out, but $x_0$ is:
$w'=x_{i_1-1} \cdots x_{i_m-1} x_{i_{m+1}} \cdots \allowbreak
x_{i_p} x_{j_n}\inv \cdots x_{j_{q+1}}\inv \bm{x_{t+m-1}\inv}
x_{j_q-1}\inv \cdots x_{j_{r+1}-1}\inv x_0^{1-r}$, where $p'=p$,
$n'=n$, $m_p'=m_p$ (because the first $m$ positive letters, whose
indices have changed, contained no bad letters), and $m_n'$ again
may only increase, if it's attained at $x_{t+m-1}\inv$. Repeating
the same calculations shows that $\dAsM(w')=\dAsM(w)$ in this case
too.

Now consider the inverse $x_tx_0\inv$ and denote $w'=x_tx_0\inv
w$. The four possible outcomes are:

\textbf{(1)} $x_0\inv$ is cancelled out, but $x_t$ is not: $x_0\inv$
can only be cancelled out if $i_1=0$, and the resulting word is:
$w'=x_{i_2} \cdots x_{i_m} \bm{x_{t+m-1}} x_{i_{m+1}} \cdots x_{i_p}
x_{j_n}\inv \cdots \allowbreak x_{j_1}\inv$. Here $p'=p$, $n'=n$,
$m_n'=m_n$ (negative part is not affected) and $m_p'=m_p$ because
the letters $x_{i_2}$ to $x_{i_m}$ cannot be bad and the relative
position of other positive letters has not changed.

\textbf{(2)} Both $x_0\inv$ and $x_t$ are cancelled out: Assuming
$x_t$ is cancelled out (due to violation of \nfii) with
$x_{j_q}\inv$, $w'=x_{i_2} \cdots x_{i_m} x_{i_{m+1}-1} \cdots
x_{i_p-1} \allowbreak x_{j_n-1}\inv \cdots x_{j_{q+1}-1}\inv
x_{j_{q-1}}\inv \cdots x_{j_1}\inv$. Here $p'=p-1$, $n'=n-1$,
$m_p'=m_p$, because $x_{i_2}$ to $x_{i_m}$ cannot be bad and the
relative position of other positive letters has not changed, and
$m_n'=m_n$, because the letters whose positions shifted also had
their indices decreased.

\textbf{(3)} Neither $x_0\inv$, nor $x_t$ are cancelled out.
$w'=x_{i_1+2} \cdots x_{i_m+2} \bm{x_{t+m}} \allowbreak
x_{i_{m+1}+1} \cdots x_{i_p+1} x_{j_n+1}\inv \cdots x_{j_q+1}\inv
\bm{x_0^{-q}}$. Here $p'=p+1$, $n'=n+1$, $m_p'=m_p$, because indices
above $i_m$ grew by 1, as did their positions, and indices
$i_1,\dots,i_m$ cannot be bad, and also $m_n'=m_n$, because all
letters whose indices increased ($j_q$ and above) shifted in
position accordingly.

\textbf{(4)} $x_0\inv$ is not cancelled out, but $x_t$ is:
$w'=x_{i_1+2} \cdots x_{i_m+2} x_{i_{m+1}} \cdots x_{i_p}
\allowbreak x_{j_n}\inv \cdots x_{j_{q+1}}\inv x_{j_{q-1}+1}\inv
\cdots x_{j_r+1}\inv \bm{x_0^{-r}}$, the cancelled pair being
$(x_{t+m},x_{j_q}\inv)$, where $j_q=t+m$. In this case, any positive
letters that can be bad kept their indices and positions, the
negative letters $j_{r+1},\dots,j_{q-1}$ had their indices and
positions shifted, while the letters $j_{q+1},\dots,j_n$ kept their
indices and positions. So $m_p'=m_p$ and $m_n'=m_n$ and obviously
$p'=p$ and $n'=n$.

We see that in all the possible cases, $\dAsM(w')=\dAsM(w)$. This
completes the proof. \qed\end{proof}
\end{clmm}

\section{Experimental results}\label{sec:distatk}

To test the applicability of the subgroup distance functions to
cryptanalysis, we tested Algorithm \ref{algo:dist} against the
Shpilrain-Ushakov protocol in the settings of Thompson's group.
Initially, each of the five distance functions presented in the
previous section was tested separately: we generated a public
element $azb$ and tried to recover a single private element $a$ or
$b$ from it. For the recovery of $a$, the functions $\dBs$ and
$\dBsW$ were used to assess the quality of the complements.
Similarly, for the recovery of $b$, we tried $\dAs$, $\dAsW$ and
$\dAsM$.

For each distance function, the experiment was run at least 1000
times, each time with new, randomly generated keys, with the minimum
recommended parameters of $s=3,L=256$. The bound $N=2L$ was chosen
on the number of iterations, since preliminary experiments have
shown that the success rates do not increase beyond that. The
results are summarized in Table \ref{tbl:dist}. It can be seen that
the distance functions $\dBs$ and $\dAsM$ noticeably outperform the
other distance functions, in recovering $a$ and $b$, respectively.
The fact that $\dAsM$ clearly outperforms its counterparts suggests
that the notion of invariance may be useful for assessing the
suitability of a given distance function.

\begin{table}
\caption{Success rates for the different subgroup distance
functions}\label{tbl:dist}
\begin{center}
\begin{tabular}{|r||r|r||r|r|r|}
\hline & $\dBs$ & $\phantom{\Big(}\dBsW$ & $\dAs$ & $\phantom{\Big(}\dAsW$ & $\dAsM$ \\
\hline
Recovery probability & 11.7\% & 3.4\% & 3.7\% & 3.4\% & 23.3\% \\
\hline
\end{tabular}
\end{center}
\end{table}

Preliminary experiments have shown that, regardless of the settings,
the success probability of finding $a_1$ given $a_1zb_1$ is similar
to that of finding $a_2\inv$ given $a_2\inv z\inv b_2\inv$. A
similar assertion holds for $b_2$ and $b_1\inv$. Therefore, in order
to estimate the overall success rate against an actual instance of
the cryptosystem, it's sufficient to try to recover one of the four
$a$'s and $b$'s. If we denote by $p_a$ and $p_b$ the probability of
successfully recovering $a$ and $b$, respectively, and assume that
all probabilities are independent, then, the expected total success
rate is roughly $1-(1-p_a)^2(1-p_b)^2$ (because each instance of the
protocol contains two elements of type $a$ and two of type $b$).

When the success rates of the two best distance functions, $\dBs$
for $a$ and $\dAsM$ for $b$, are combined, the expected overall
success probability, according to the above, is between 50\% and
54\%, which was experimentally verified. Note that this attack is
very efficient, since it involves no backtracking, no lookahead, and
no analysis of suboptimal partial results: it tries to peel off the
generators by a greedy algorithm, which considers only locally
optimal steps. Attacking each key required only a few seconds on a
single PC, and it is very surprising that such a simple attack
succeeds about half the time. These results are much better than
those achieved by length-based attacks of similar complexity on this
cryptosystem (see \cite{Thomp1}).

It is interesting to note that possible extensions of the attack,
such as memorizing many suboptimal partial solutions or using
significant lookahead (which require much higher time and space
complexities) have different effects on length-based and
distance-based attacks. While it was shown in \cite{Thomp1} that these
extensions greatly improve the success rates of the length-based
attack, experiments with the distance-based attack, with similar
values of the memory and lookahead parameters, showed almost no
improvement. However, the situation may be very different for other
cryptosystems and other subgroup distance functions.

To further test the performance of the distance functions, several
experiments were run with different values of the parameters
$(s,L)$. We used the combination of $\dBs$ and $\dAsM$, which was
established as the best in the former experiment. Table
\ref{tbl:distvpar} shows the overall success probability, for
$L\in\{128,256,320,512,640,960\}$ and $s\in\{3,5,8\}$. The success
rates stay remarkably consistent across different lengths for a
given $s$, and even increasing $s$ does not cause a significant
drop. The time complexity of the attack grows linearly with $s$ and
roughly quadratically with $L$, with most of the time being spent on
computing normal forms of elements in the group. For the largest
parameters presented here, the attack still required under a minute
in most cases. This suggests that for the Shpilrain-Ushakov
cryptosystem the distance-based attack remains a viable threat, even
when the security parameters $s$ and $L$ are increased beyond the
original recommendations.

\begin{table}
\caption{Success rates for different combinations of
$(s,L)$}\label{tbl:distvpar}
\begin{center}
\begin{tabular}{|r|r|r|r|r|r|r|}
\hline
& $L=128$ & $L=256$ & $L=320$ & $L=512$ & $L=640$ & $L=960$ \\
\hline
$s=3$ & 51.7\% & 47.9\% & 55.5\% & 51.2\% & 50.4\% & 52.6\% \\
\hline
$s=5$ & 46.0\% & 47.1\% & 48.4\% & 51.1\% & 48.2\% & 48.3\% \\
\hline
$s=8$ & 36.2\% & 42.8\% & 41.3\% & 46.5\% & 42.4\% & 50.3\% \\
\hline
\end{tabular}
\end{center}
\end{table}

\section{Conclusion}

We introduced a novel form of heuristic attacks on public key
cryptosystems that are based on combinatorial group theory, using
functions that estimate the distance of group elements to a given
subgroup. Our results demonstrate that these distance-based attacks
can achieve significantly better success rates than previously
suggested length-based attacks of similar complexity, and thus they
are a potential threat to any cryptosystem based on equations in a
noncommutative group, which takes its elements from specific
subgroups. It will be interesting to test this approach for other
groups and other protocols.


\begin{thebibliography}{99}

\bibitem{AAG}
I.\ Anshel, M.\ Anshel and D.\ Goldfeld, \emph{An algebraic method
for public-key cryptography}, Mathematical Research Letters
\textbf{6} (1999), 287--291.

\bibitem{Artin}
E.\ Artin, \emph{Theory of Braids}, Annals of Mathematics
\textbf{48} (1947), 127--136.

\bibitem{IntroThompson}
J.W.\ Cannon, W.J.\ Floyd and W.R.\ Parry, \emph{Introductory notes
on Richard Thompson's groups}, L'Enseignement Mathematique (2)
\textbf{42} (1996), 215--256.

\bibitem{Ts1}
D.\ Garber, S.\ Kaplan, M.\ Teicher, B.\ Tsaban, and U.\ Vishne,
\emph{Length-based conjugacy search in the Braid group},
Contemporary Mathematics \textbf{418} (2006), 75--87.

\bibitem{Ts2}
D.\ Garber, S.\ Kaplan, M.\ Teicher, B.\ Tsaban, and U.\ Vishne,
\emph{Probabilistic solutions of equations in the braid group},
Advances in Applied Mathematics \textbf{35} (2005), 323--334.

\bibitem{HugTan} J.\ Hughes and A.\ Tannenbaum,
\emph{Length-based attacks for certain group based encryption
rewriting systems}, Workshop SECI02 S\'ecurit\'e de la Communication
sur Internet (2002).

\bibitem{KoLee}
K.H.\ Ko, S.J.\ Lee, J.H.\ Cheon, J.W.\ Han, J.\ Kang and C.\ Park,
\emph{New Public-Key Cryptosystem Using Braid Groups}, Lecture Notes
in Computer Science \textbf{1880} (2000), 166--183.

\bibitem{Matucci}
F.\ Matucci, \emph{The Shpilrain-Ushakov Protocol for Thompson's
Group $F$ is always breakable}, e-print
\texttt{arxiv.org/math/0607184} (2006).

\bibitem{Thomp1}
D.\ Ruinskiy, A.\ Shamir, and B.\ Tsaban,
\emph{Length-based cryptanalysis: The case of Thompson's Group},
Journal of Mathematical Cryptology \textbf{1} (2007), 359--372.

\bibitem{ShpAssessing}
V.\ Shpilrain, \emph{Assessing security of some group based
cryptosystems}, Contemporary Mathematics \textbf{360} (2004),
167--177.

\bibitem{Unnec}
V.\ Shpilrain and A.\ Ushakov, \emph{The conjugacy search problem in
public key cryptography: unnecessary and insufficient}, Applicable
Algebra in Engineering, Communication and Computing \textbf{17}
(2006), 291--302.

\bibitem{ShpUsh}
V.\ Shpilrain and A.\ Ushakov, \emph{Thompson's group and public key
cryptography}, ACNS 2005, Lecture Notes in Computer Science
\textbf{3531} (2005), 151--164.

\end{thebibliography}
\end{document}